\documentclass[tightenlines,aps,floatfix,preprint,nofootinbib]{revtex4}

\usepackage{float,epsfig}

\newcommand{\ba}{\begin{eqnarray}}
\newcommand{\ea}{\end{eqnarray}}

\newcommand{\be}{\begin{equation}}
\newcommand{\ee}{\end{equation}}

\begin{document}

\hfill$\vcenter{
\hbox{\bf FERMILAB-PUB-06-054-T}
\hbox{\bf MADPH-06-1254}
\hbox{\bf UH-511-1084-06}}
$
\vspace{1cm}

\title{
Combining Monte Carlo generators with next-to-next-to-leading order calculations: event reweighting for 
Higgs boson production at the LHC\\ \vspace{0.25cm}
}

\author{
Giovanna Davatz$^a$,
Fabian St\"ockli$^a$,
Charalampos Anastasiou$^a$, 
G\"unther Dissertori$^a$,
Michael Dittmar$^a$,
Kirill Melnikov$^b$,
Frank Petriello$^c$\\
}

\affiliation{
$^a$Institute of Particle Physics, ETH, 8093 Zurich, Switzerland\\
$^b$Department of Physics and Astronomy, University of Hawaii, 2505 Correa Rd., Honolulu, Hawaii 96822\\
$^c$University of Wisconsin, Madison, WI 53706, USA and
Fermi National Accelerator Laboratory, P.O. Box 500, MS 106, Batavia, IL 60510,
USA}

\begin{abstract}

We study a phenomenological ansatz for merging next-to-next-to-leading order (NNLO) calculations 
with Monte Carlo event generators.  We reweight them to match bin-integrated NNLO differential distributions.  
To test this procedure, we study the Higgs boson production cross-section at the LHC, for which a fully differential 
partonic NNLO calculation is available.
We normalize PYTHIA and MC@NLO Monte Carlo events for Higgs production in the gluon fusion channel to reproduce the bin integrated 
NNLO double differential distribution in the
transverse momentum and rapidity of the Higgs boson. 
These events are used to compute differential distributions for the photons in the 
$pp \to H \to \gamma \gamma$ decay channel, and are compared to predictions from 
fixed-order perturbation theory at NNLO. 
We find agreement between the reweighted generators and the NNLO result 
in kinematic regions where we expect a good description using fixed-order perturbation theory. 
Kinematic boundaries where resummation is required are also modeled correctly using this procedure.
We then use these events to compute distributions in the 
$pp \to H \to W^+W^- \to l^+l^- \nu\bar{\nu}$ channel, for which an accurate description 
is needed for measurements at the LHC.  We find that the final state lepton distributions obtained from PYTHIA are 
not significantly changed by the reweighting procedure.

\end{abstract}

\maketitle


\section{Introduction}
The search for the Higgs boson is a main 
objective of the LHC physics program. The ATLAS and CMS detectors
are designed to detect a Higgs boson in the mass range 
from about 100 GeV up to at least 600 GeV.
During the last 15 years, many Higgs boson signatures have been studied. 
For a detailed description of this effort we refer 
the reader to Refs.~\cite{search,jakobs,djouadi_rev,carena}. 
If ATLAS and CMS function as designed, a discovery 
of a Standard Model Higgs boson over the entire mass range can be expected with luminosities of 
about 30 fb$^{-1}$. 

It is interesting to study how well the mass, width, and couplings of 
a particle with the properties of a Higgs boson can be measured at the LHC, 
and how well these measurements can discriminate 
between the Standard Model and its viable extensions. 
The potential statistical accuracy for such 
measurements is usually assessed by computing experimental 
efficiencies using leading order (LO) parton shower  
Monte Carlo generators. However, these efficiency estimates cannot be expected 
to be accurate, especially in complicated signatures such as 
$pp \to H \to W^+ W^-$. 

It has been found  that corrections beyond LO are  particularly significant 
for Higgs boson production in the channel $gg \to H$. The 
next-to-leading-order (NLO) QCD corrections~\cite{dawson,spira} 
increase the cross-section by a factor of about 1.7-2 as compared 
to the LO result.
A few years ago, the inclusive cross-section was computed in the large 
$m_{top}$ limit with NNLO accuracy~\cite{hnnlo_harl, hnnlo_an, hnnlo_rav}. 
The new corrections increase the NLO cross-section by another $20-25\%$.  
Very recently, some threshold-enhanced ${\rm N^3LO}$ terms were also 
computed~\cite{hnnnlo}, changing the NNLO result by less than $5 \%$. 
These computations show that the Higgs boson cross-section can 
be reliably estimated only after  many orders in perturbation theory have 
been considered. It is therefore  necessary to account for 
higher order corrections  in a realistic  analysis of the Higgs boson 
signal.

Recently, a 
program FEHIP that describes
Higgs boson production
in gluon fusion  was developed~\cite{hnnlo_diff,fehip}. 
FEHIP  computes the cross-section and fully differential 
distributions for Higgs boson production at NNLO in QCD. 
Arbitrary 
cuts can be imposed on partonic 
jets and on the decay products of the Higgs boson.  
FEHIP does not have an implementation of a parton shower and a hadronization 
algorithm. This creates a few shortcomings, since  it is not possible to apply 
cuts at the hadron level or to generate events for a detector simulation. 
In addition, regions of phase-space close to kinematic 
boundaries can not be described  
reliably in fixed-order calculations. This feature manifests itself 
in  large perturbative corrections at special kinematic regions, 
such as the low Higgs $p_\bot$ region. 

To overcome these characteristic problems of fixed order 
perturbative computations, the resummation of soft gluon effects 
to all orders in perturbation theory must be performed.  This 
resummation may be obtained using analytical techniques~\cite{resbos,respt}.  
Alternatively, these effects are also included in parton shower Monte Carlo programs.  
Novel approaches~\cite{mcatnlo,nason,nagy,soper} 
merge cross-sections  computed in 
fixed order perturbation theory with these LO 
event generators, such as PYTHIA~\cite{pythia} and HERWIG~\cite{herwig}.
Very significant progress has been achieved, and 
the pioneering Monte-Carlo event generator MC@NLO~\cite{mcatnlo} 
combines consistently NLO perturbative calculations with HERWIG 
for a number of processes at  hadron colliders. 
Unfortunately, no  method exists which merges parton shower 
algorithms with NNLO partonic cross-sections consistently.  
This is desirable for processes with large 
perturbative corrections.

It is possible to incorporate NNLO corrections into realistic 
analyses of experimental signatures in an approximate way 
by multiplying probabilities of events in a parton shower Monte Carlo simulation 
by so-called $K$-factors.
These factors force the Monte Carlo output to agree with 
certain observables computed in perturbative QCD.  
This technique is called event reweighting. The simplest version  of this 
technique is a multiplication of the Monte Carlo output by a constant 
factor so that the total cross-sections computed perturbatively and 
with the reweighted Monte-Carlo simulation agree.

Re-scaling Monte Carlo output by a constant factor 
does not guarantee an agreement between perturbative and Monte Carlo
results for differential distributions, since perturbative corrections 
do depend on kinematic variables and vary across the phase-space.
A better job may be done if the Monte Carlo output 
and the perturbative calculation are matched at 
the differential level~\cite{reweight}. 
A point-by-point reweighting of the Monte-Carlo throughout the available phase-space 
is not possible, since infrared divergences would produce divergent weights.
We must instead select a realistic set of observables to match,
 and then check if the reweighted simulation
gives a reliable prediction for other observables.

In this paper we study the reweighting procedure for Higgs boson production 
in the gluon fusion channel at the LHC.
We match the Monte-Carlo output of both PYTHIA and MC@NLO to distributions 
that depend only on the Higgs boson kinematics, which is a simple and obvious 
way of reweighting the Monte Carlo output.
We match to a double differential distribution in the Higgs
boson transverse momentum and rapidity.  This distribution is chosen both for its 
simplicity and because it allows us to decouple the Higgs boson decay 
chain from the Higgs boson production.
However, the kinematics of accompanying 
QCD radiation is totally ignored in the reweighting process. 
This ignorance is not a problem if hadronic radiation 
is treated fairly inclusively by cuts applied to a process of interest; 
however, if a detailed description of the hadronic radiation becomes 
relevant, the reweighting procedure may lead to inaccurate results.
A particular example of a situation when this 
happens is the jet veto on transverse 
momenta of hadronic jets; we discuss it in detail in 
Section~\ref{sec:FEHIP}.

The paper is  organized as follows.
In  Section~\ref{sec:FEHIP} we introduce the reweighting procedure 
and discuss in detail differential distributions 
in the reaction $p p \to H + X$.  We first study the reweighting procedure 
at NLO by comparing the fixed order result with PYTHIA and MC@NLO.  We present an 
example in which the reweighting procedure fails to produce accurate acceptances: 
when a jet veto is imposed on the transverse momenta of extra QCD radiation.  We explain 
how this problem is ameliorated at NNLO.

We then apply the reweighting approach
to estimate the NNLO effects 
for the channels $pp \to H \to \gamma \gamma $
and $pp \to H \to W^+W^- \to l^+ \nu l^- \bar{\nu}$. 
We first reweight PYTHIA and MC@NLO events in 
the $pp \to H \to \gamma \gamma$ channel.  
We compute the accepted cross-section and differential 
distributions which have a potential discriminating power from the 
di-photon irreducible background~\cite{lance1,fehip,fabian}: the 
average $p_m = 
(p_\bot^{\gamma_1}+p_\bot^{\gamma_2})/2$ transverse momentum distribution and the 
pseudorapidity difference $\eta^* = \left|\eta^{\gamma_1}-\eta^{\gamma_2}\right|/2 $ of the two photons. 
We find an excellent agreement between the reweighted 
PYTHIA and MC@NLO events for  all observables. The di-photon channel is a 
testing ground for the reweighting procedure, since we can compare the results 
with the NNLO predictions of FEHIP for the same observables. 
We find that  accepted cross-sections agree better than $1\%$. 
The $p_m$, 
$\eta^*$ distributions also agree very well away from kinematic thresholds.  
Near these boundaries, they reproduce the correct resummed behavior of the parton-shower
Monte Carlo simulations. 

We next study the $pp \to H \to W^+W^- \to l^+l^-\nu \bar{\nu}$ channel. 
Ref.~\cite{reweight} already employed  a reweighting technique in order to 
study the effect of perturbative corrections in this channel 
by matching the PYTHIA output to the resummed $p_\perp$ spectrum of the 
Higgs boson \cite{respt}.
An optimal set of cuts for isolating  a Higgs signal in this channel 
was introduced and studied in~\cite{Dittmar:1996ss}.  
A study of this channel with higher-order QCD corrections included 
was presented in~\cite{newCMS}, where both the signal and $qq \rightarrow WW$ background 
were included using a reweighting technique.  The analysis in Section~\ref{sec:FEHIP} of the accuracy 
achievable by reweighting PYTHIA for processes with a jet veto cut implies 
that at least {\it shapes} of distributions can be predicted reliably.
Hence, we calculate  various lepton distributions in the reaction 
$pp \to H \to W^+W^- \to l^+ l^- \nu \bar \nu$.  Interestingly, 
the reweighting turns out to be largely irrelevant for these distributions 
and the prediction of reweighted PYTHIA and standard PYTHIA agree very well.

\section{The reweighting technique for  $pp \to H +X$}
\label{sec:FEHIP}

\subsection{The reweighting procedure}

 The cross-sections computed with  generator 
$G = \{{\rm PYTHIA, \; MC@NLO}\}$ for the process
$pp \to H+X$ are
\begin{equation}
\label{eq:eg_xsection}
\sigma^{G} = \sum_m \int d\Pi_m f_m^{G}\left(\left\{p_i\right\}\right) 
{\cal O}_m\left(\left\{p_i\right\}\right), 
\end{equation}
where we sum over all final-state multiplicities $m$, and integrate the events
$f_{m}^{G}$ over the phase-space variables $d\Pi_m$ of all $i \leq m$
particles in the final state. 
The function ${\cal O}_m$ selects the 
kinematic configurations to be accepted in the 
measured cross-section. The events depend implicitly on the 
renormalization and various factorization scales.

The simplest observable is the total cross-section $\sigma^{G}_{\rm incl}$, 
corresponding 
to ${\cal O}_m\left(\left\{p_i\right\}\right)=1$. It is a well-known fact  
that standard event generators fail to predict total cross-sections reliably.
As an example we set the mass of the Higgs boson to $m_H = 165~{\rm GeV}$, 
and the renormalization and factorization scales to 
$\mu_R=\mu_F=m_H/2$.  We use the generators PYTHIA version 6.325 with a $Q^2$ ordered 
parton shower and MC@NLO version 3.2
For PYTHIA we use the MRST2001 LO set of 
parton-distribution functions, while  for MC@NLO we use the corresponding 
NLO set.  In MC@NLO, the scale choice $\mu_R=\mu_F=m_H/2$ is also used, 
instead of the default setting.  The resulting PYTHIA and MC@NLO cross-sections are
\begin{equation} 
\sigma_{\rm incl}^{\rm PYTHIA}  = 12.20 \,{\rm pb} \quad \quad
\sigma_{\rm incl}^{\rm MC@NLO}  = 23.92 \,{\rm pb}.
\label{eq1}
\end{equation}
The corresponding fixed-order NNLO cross-section is
 \begin{equation} 
\sigma_{\rm incl}^{\rm NNLO}  = 27.78 {\rm pb}.
\label{eq2}
\end{equation}
The large differences between the PYTHIA, MC@NLO and NNLO 
cross-sections  reflect the fact that the NLO and 
NNLO perturbative corrections are very significant. 

A consistent method  for merging fixed-order perturbative calculations  
and parton-shower algorithms is only formulated at NLO in perturbation 
theory, and is implemented in MC@NLO. A similar procedure beyond NLO
is not yet available. Nevertheless, we would
like to incorporate the large perturbative corrections into the 
event generators. 
In this paper, we adopt a pragmatic approach
 to solve this problem. We multiply 
the integrand in Eq.(\ref{eq:eg_xsection}) with a function $K^{G}$, 
\begin{equation}
\label{eq:reg_xsection}
\sigma^{R(G)} = \sum_m \int d\Pi_m f_m^{G}\left(\left\{p_i\right\}\right)
K^{G}\left(\left\{p_i\right\}\right) 
{\cal O}_m\left(\left\{p_i\right\}\right), 
\end{equation}
in order to reweight the events $f_{m}^{G}$, 
\[
f_m^{G} \to f_m^{R(G)} = f_m^{G} K^{G}.  
\]
The reweighting factors $K^{G}$ 
model the effect of higher order corrections through a certain 
order in perturbation theory.
We determine 
the factors $K^{G}$ by requiring that
Eq.(\ref{eq:reg_xsection}) 
reproduces the fixed-order perturbative results  for 
selected distributions,
\begin{equation}
\sigma^{R(G)}\left( {\cal O}_{\rm special} \right)
=\sigma^{\rm PT}\left( {\cal O}_{\rm special} \right). 
\end{equation}
We emphasize that Eq.(\ref{eq:reg_xsection}) 
is an approximate 
ansatz to describe effects of higher order corrections in the 
absence of a rigorous treatment. 
Strictly speaking, higher order corrections do depend on parton multiplicities. 
For example, $pp \to H + 0~{\rm jets}$ is renormalized 
differently compared to, $pp \to H + 1~{\rm jet}$. This feature 
is ignored in Eq.(\ref{eq:reg_xsection}), where the reweighting factors 
$K_G$ do not depend on the multiplicities $m$. 
A more detailed version of reweighting would 
not be universal, because 
matrix elements with fixed multiplicities 
of partons are divergent in perturbation theory. 
Independent 
renormalization of events with different multiplicities has to depend 
on a globally-defined set of cuts, e.g. the jet finding algorithm. 
This invalidates the unweightedness of events, 
the single most important feature of parton shower Monte Carlo event generators.
We will see the errors in the reweighting procedure 
caused by neglecting the dependence on parton multiplicities later in this 
section, when we compare reweighted PYTHIA at NLO with MC@NLO.

Having pointed out the approximate nature of the reweighting procedure, 
we discuss a choice of a suitable distribution for which the agreement 
of a Monte Carlo generator and the perturbative calculation can be imposed.
Since, as we discussed in the previous paragraph, the reweighting ansatz 
is unsuitable for resolving the structure of QCD radiation, we use the 
kinematic variables which describe the Higgs boson.
Since,
 up to an angle in a plane transverse to the collision axis, 
the Higgs boson kinematics  is  determined by its transverse 
momentum $p_\bot$ and  rapidity $Y$,
we  normalize the events $f_{m}^{G}$ to the magnitudes and shapes of the 
NNLO bin-integrated double 
differential distributions in $Y$ and $p_\bot$. 
We expect that such a normalization
 renders the events
more realistic in predicting other observables of the process.  Without 
a technique for combining NNLO results with parton showering in the 
spirit of MC@NLO, this is the best way we have of combining these calculations 
with event generators.
Note that we are not changing the properties of the radiation produced 
by the Monte Carlo generators.  We are only changing the normalization of 
these events to reproduce certain distributions.  The reweighted generators 
therefore do not better describe events with multiple hard radiations.  

We choose  
\begin{equation}
\label{eq:special_obs}
{\cal O}_{\rm special} =\left\{ \begin{array}{l} 
        1, \mbox{ if } p_\bot \in \left[p_\bot^{j}, p_\bot^{j+1}\right] 
\mbox{ and }  Y \in \left[Y^{i}, Y^{i+1}\right] \\
       0, \mbox{ otherwise, } 
     \end{array} \right \}
\end{equation}
and define the $K$-factors as 
\begin{equation}
K^{G}_{ij}:=K^{G}(\{ p_f \}) = 
\frac{\Delta \sigma_{ij}^{\rm PT}}{\Delta \sigma_{ij}^{G}} \quad 
\mbox{ if } p_\bot \in \left[p_\bot^{j}, p_\bot^{j+1}\right] 
\mbox{ and }  Y \in \left[Y^{i}, Y^{i+1}\right],
\end{equation}
where $\Delta \sigma_{ij}^{{\rm PT},G}$ are the accepted cross-section 
computed at fixed order perturbation theory and with the generator $G$,
respectively.
The values of the bin boundaries $p_\bot^j$ and $Y^i$ are chosen in 
such a way that they 
capture the shape of the Higgs  
$p_\bot$ and rapidity distributions and span the
allowed kinematic range for $Y$ and $p_\bot$. 
In what follows we always set the renormalization and factorization scales 
to $\mu_R=\mu_F=m_H/2$, since this choice is known to yield a 
perturbative series with faster convergence~\cite{hnnlo_an}.

\begin{figure}[htbp!]
\begin{center}
\includegraphics[width = 8cm,angle=0]{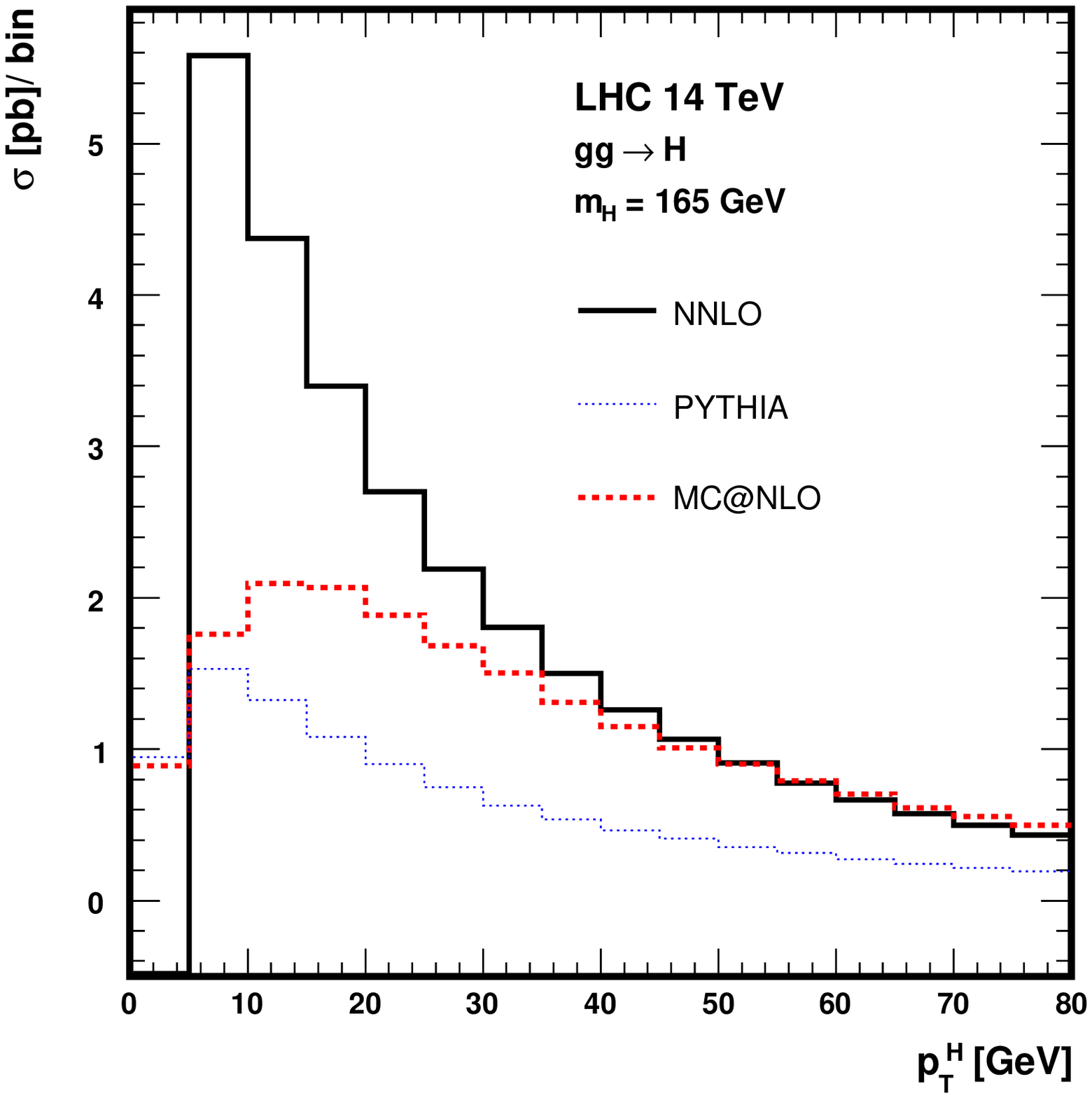}
\includegraphics[width = 8cm,angle=0]{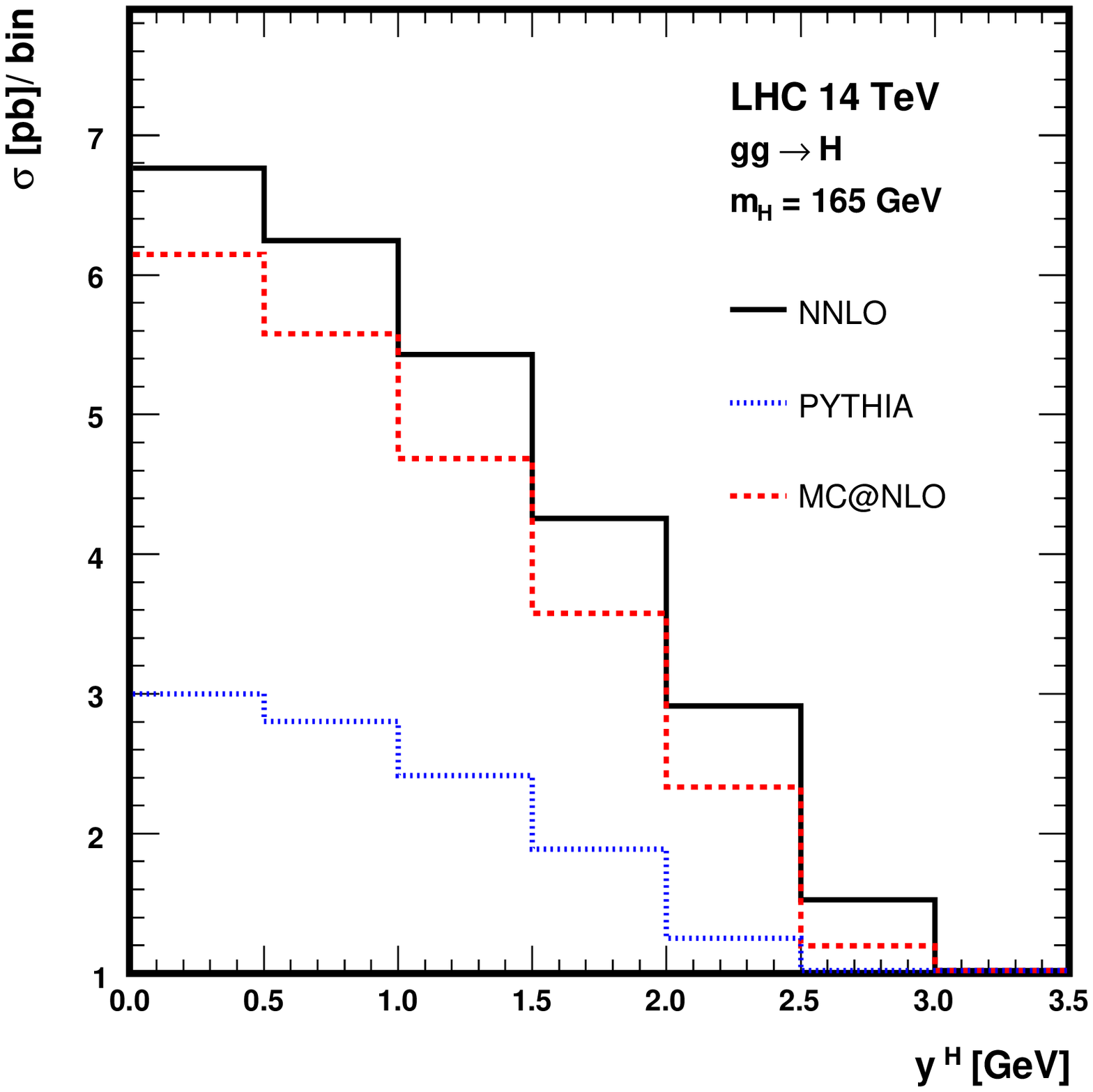}
\end{center}
\caption{\label{fig:bin0a} 
Distributions of the Higgs transverse momentum (left) and rapidity (right).}
\end{figure}

We point out that 
at NLO the $p_\bot$ and $Y$  variables completely 
constrain the kinematics of the process 
$pp \to H +X$. At NNLO, new  tree level  
processes $pp \to H + i+j$ with two partons in the final
 state require additional 
phase-space variables in order to determine the kinematics of  partonic 
radiation.
Our reweighting functions $K^{G}$ are independent of 
these additional variables.  This 
may create a problem in case there are important cuts on the hadronic 
radiation.


The choice of the bin boundaries in $p_\bot$ of the Higgs boson   
is a subtle issue.
Some of the standard cuts that we need to  apply in  Higgs boson 
production processes 
have a strong dependence on the $p_\bot$ of the Higgs boson. For example, 
in the $WW$ decay channel it is important to have a phenomenologically 
realistic model for the low and intermediate $p_\bot$ region since 
after cuts most of the signal comes from the region 
of low Higgs $p_\bot$.
The Higgs $p_\bot$ distribution in this region 
is not correctly described in fixed-order calculations. 
Logarithms of the form  
$\log p_\bot/m_H $ become large and require a resummation.
Nevertheless, fixed order calculations for cross-sections
integrated over $p_\bot$ of the Higgs boson are still viable, provided 
that the integration region is sufficiently broad.

In Fig.~\ref{fig:bin0a} we show the $p_\bot$ distributions 
for the fixed-order NNLO calculation, PYTHIA and MC@NLO. 
We  observe that the perturbative NNLO 
result breaks down at small $p_\bot$. 
The $p_\bot$ spectrum of PYTHIA is peaked at lower $p_\bot$ than MC@NLO.  
The most reliable spectrum at low $p_\bot$ is obtained with resummation~\cite{respt}.
To avoid  problems associated with the low-$p_\bot$ region in fixed order 
perturbative calculations, we  choose the first $p_\bot$ bin, 
$[p_\bot^0=0, p_\bot^1]$, in Eq.(\ref{eq:special_obs}) to be 
sufficiently broad by taking  $p_\bot^1 = 25$ GeV.
Therefore, for $p_\bot< 25{\rm \, GeV}$, 
we reweight all events with a uniform factor, 
maintaining the shape of the $p_\bot$ distribution provided 
by  the generator $G$. 
Above $25 {\rm \, GeV}$, we trust the shape of the perturbative 
result and  reweight in bins of $5 {\rm \, GeV}$
\[
p_\bot^0=0, \quad {p_\bot^1=25\,{\rm GeV}},
\quad  
p_\bot^i = (25 + (i-1)5)\, {\rm GeV},    
\]
and 
\[
Y^j = 0.5 (j-1)\, , \,j=1 \ldots 9.
\]
Note that this reweighting procedure 
leads to a discontinuity at $p_\bot = p_\bot^{1}$ 
in the reweighted  $p_\bot$ spectrum computed with the generator $G$. 
The choice of $p_\bot^{1}$ is ambiguous; however, it turns 
out that this ambiguity  is largely irrelevant in practice.
In what follows we take the first bin in $p_\bot$ to be 
$[0-25~{\rm GeV}]$, unless explicitly stated otherwise.

\begin{figure}[htbp!]
\begin{center}
\includegraphics[width =7cm,angle=0]{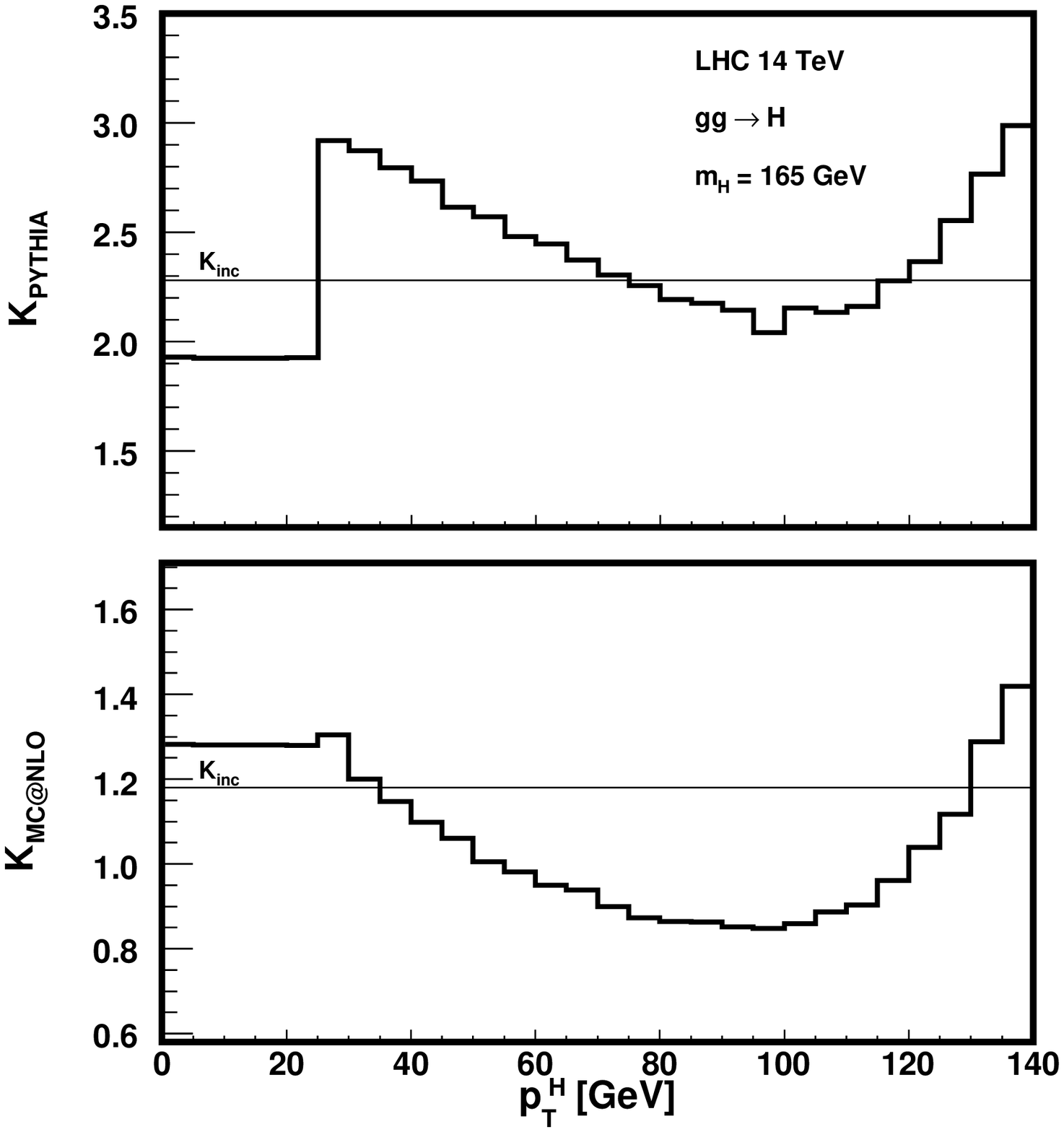}
\includegraphics[width =7cm,angle=0]{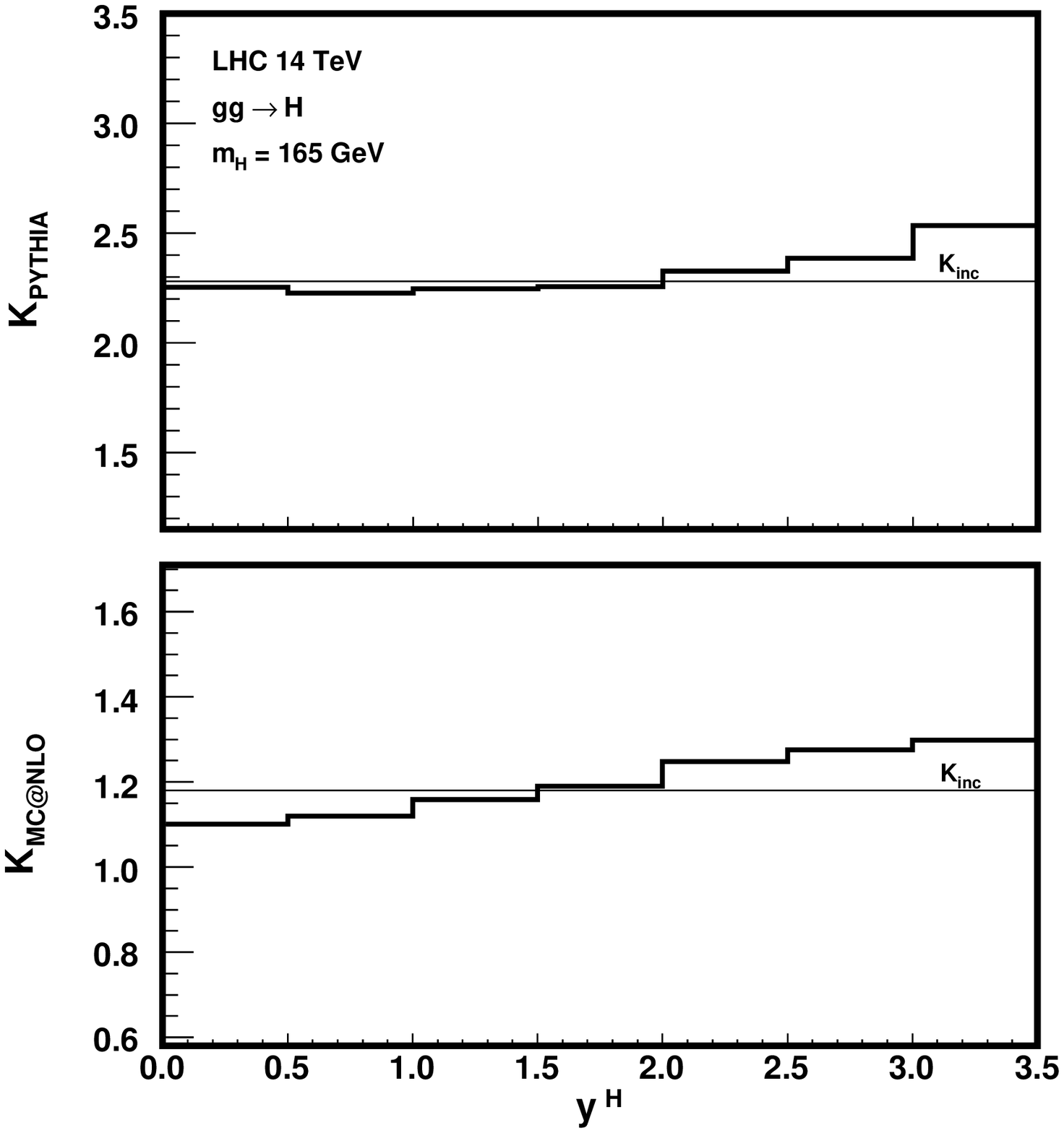}
\end{center}
\caption{\label{fig:kfac_pt_all}
The reweighting factors integrated over rapidity
for PYTHIA and MC@NLO as functions of the Higgs $p_\bot$ (left).
The reweighting factors, integrated over $p_\bot$, as
functions of rapidity (right). The inclusive $K$-factor for the total
cross-section is also shown on both plots as a horizontal line.}
\end{figure}

At this point it is worth investigating if just a single, constant $K$-factor
is sufficient for accurate reweighting. To do so, 
we investigate the dependence of the reweighting factors $K^{G}$
in Eq.(\ref{eq:reg_xsection}) on the $p_\bot$ and rapidity and 
 find that 
$K^{G}(p_\bot, Y)$ can vary significantly in different 
rapidity and $p_\bot$-bins.
For PYTHIA, we find $K$-factors ranging from 1.8 to 3.5, 
while for MC@NLO the $K$-factors can vary 
from 0.7 to 1.6 in bins with a significant number of events.   
For illustration, in Fig.~\ref{fig:kfac_pt_all} we show 
the reweighting factors  for the $p_\bot$ distribution, 
after we integrate over rapidity.  We also show the reweighting factors as a 
function of $Y$, after integrating over $p_\bot$. 
The shape of the $K$-factors in the two variables is not uniform, indicating 
that a naive multiplication with a uniform $K$-factor from the total 
cross-section may  not be adequate. Having discussed the reweighting 
technique in general, we now study it in the process $pp \rightarrow H+X$ at both NLO and NNLO.

\subsection{The reweighting at NLO and NNLO}

We now apply the reweighting procedure to PYTHIA and MC@NLO to study $pp \rightarrow H+X$.
MC@NLO is a Monte Carlo event generator 
that accommodates NLO perturbative calculations for a wide range 
of processes. Its important feature is that the parton shower 
is combined with NLO matrix elements consistently, without 
double counting.  PYTHIA is an event generator 
based on leading order matrix elements,
so that hadronic radiation is generated primarily through 
the parton shower. 

We begin our study by checking how well reweighting works at NLO.  We check how well the 
procedure describes Higgs boson production when a jet veto is imposed.  This tests whether the 
neglect of the additional hadronic radiation in our reweighting ansatz is problematic.  Phenomenologically, 
this cut is needed in the $pp \rightarrow H \rightarrow W^+W^- \rightarrow l^-l^+ \bar{\nu}\nu$ channel 
to isolate the signal from background.  The inclusive 
cross sections for both PYTHIA and MC@NLO are given in Eq.(\ref{eq1}), while 
$\sigma_{\rm incl}^{\rm NLO}=23.99$ pb.  We impose a jet veto of $p_{\perp}<30$ GeV, and we define 
jets with a cone algorithm using a cone size $R=0.4$.  
The cross sections after the jet veto has been imposed are 
\begin{equation}
\sigma_{\rm acc}({\rm pb}) =
\left \{
\begin{array}{cl}
6.12, &\;\;\;{\rm PYTHIA};\\
12.09, &\;\;\;{\rm MC@NLO};\\
14.48, &\;\;\;{\rm R^{NLO}(PYTHIA)};\\
16.34, &\;\;\;{\rm NLO}.
\end{array}
\right.
\label{NLOcr}
\end{equation}
The acceptances, defined as the ratios of the accepted cross sections over the inclusive cross sections, are 
\begin{equation}
A =
\left \{
\begin{array}{cl}
0.50,  &\;\;\;{\rm PYTHIA};\\
0.51, &\;\;\;{\rm MC@NLO};\\
0.60, &\;\;\;{\rm R^{NLO}(PYTHIA)};\\
0.68, &\;\;\;{\rm NLO}.
\end{array}   
\right.
\label{NLOacc}
\end{equation}

We observe a very large disagreement, of order 30\%, between the acceptances obtained using the generators and the fixed order 
NLO result.  What is occurring here is that this observable is very sensitive to the properties of the QCD radiation.  
Multiple partonic emissions are required to generate the correct jet $p_{\perp}$ spectrum, and the NLO result 
contains only a single partonic emission.  The $p_{\perp}$ spectrum of this additional parton is generated for the 
first time at NLO, and is therefore not accurately predicted at this order in the perturbative expansion.  
We note that reweighting PYTHIA to the NLO result spoils the agreement between its acceptance and that computed with MC@NLO.  

\begin{figure}[h]
\begin{center}
\includegraphics[width = 8cm,angle=0]{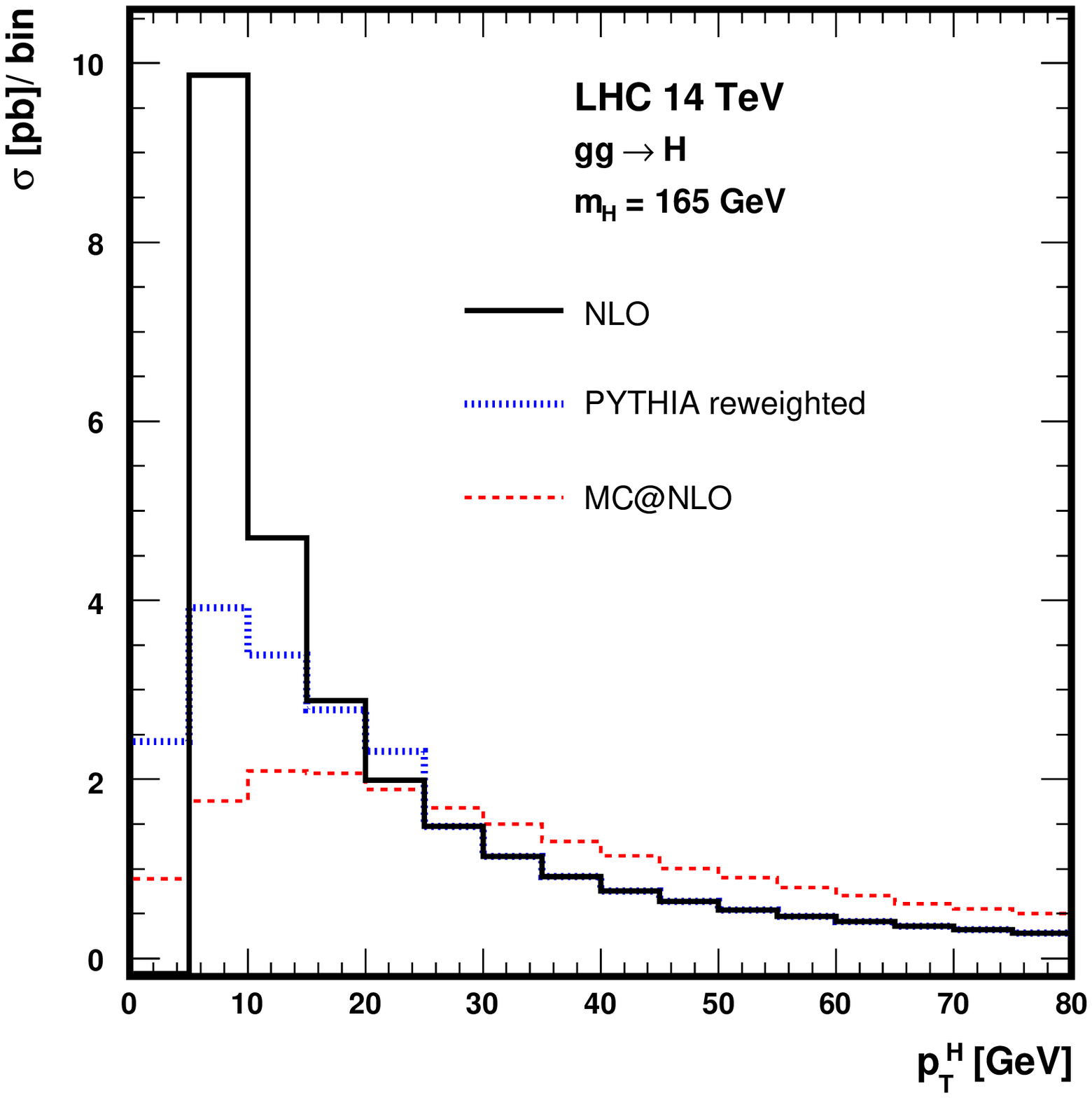}
\includegraphics[width = 8cm,angle=0]{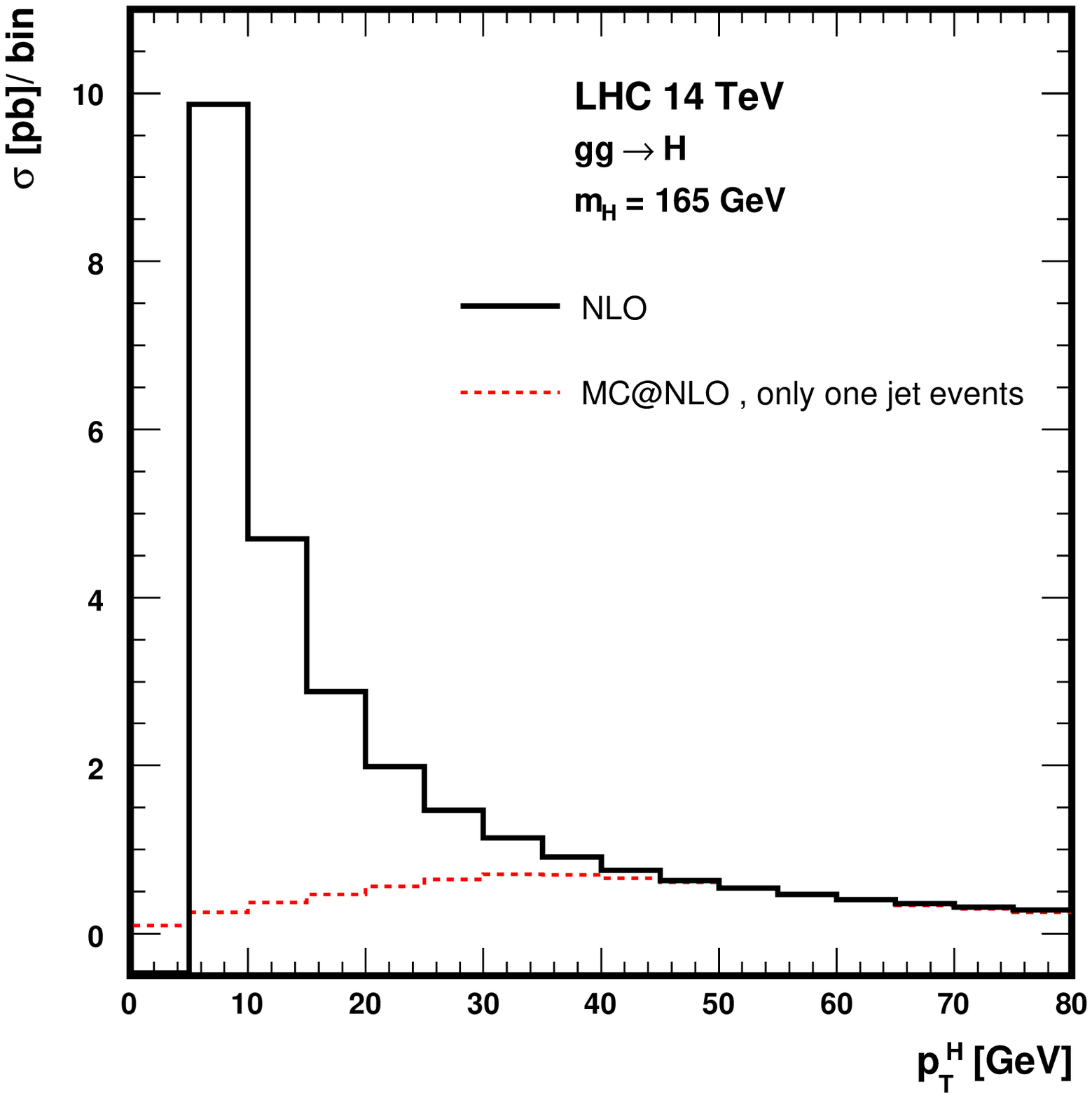}
\end{center}
\caption{\label{fig:NLOpt} 
The Higgs boson $p_{\perp}$ spectrum for NLO, MC@NLO, and PYTHIA reweighted (left panel); 
comparison of the Higgs boson $p_{\perp}$ at NLO, and with MC@NLO when only a single jet is observed (right panel).}
\end{figure}

To check that multiple emissions are indeed important, we present below the jet multiplicites for both 
PYTHIA and MC@NLO before a jet veto is imposed.  We study the cross-section both inclusively and with the restriction 
$p^{H}_{\perp}>30$ GeV, to show that multiple emissions are required to obtain 
correctly even the high $p_{\perp}$ events.  We present the fraction of events with $N=0,1,2,3,4$, or more jets in 
Table~\ref{NLOtab}.  Note that we require a jet to have $p_{\perp}> 20$ GeV, so events without jets are possible.  
Over half of the events in the high 
$p^{H}_{\perp}$ tail coming from PYTHIA and MC@NLO contain multiple emissions,
indicating that the description of the hadronic radiation coming from the single emission at NLO is unlikely
to be very accurate.  We can also see this by studying the Higgs $p_{\perp}$ spectrum, shown in the left panel of Fig.~\ref{fig:NLOpt}.
The single hard partonic emission is equivalent in the fixed order NLO result and in MC@NLO.
The mismatch between them in the high $p^{H}_{\perp}$ tail is caused by showering.
The importance of the multiple emissions is made explicit in the right panel of Fig.~\ref{fig:NLOpt}, where the $p^{H}_{\perp}$
from MC@NLO when only a single jet is observed is compared to the NLO calculation.  The distributions agree very well in the
high $p^{H}_{\perp}$ region for this single emission case, again indicating the need for multiple emissions to
correctly generate this spectrum.

\begin{tiny}
\begin{table}[h]
\begin{center}
\begin{tabular}{||c||c|c||c|c||} \hline
& \multicolumn{2}{c||}{\emph{Inclusive}} & \multicolumn{2}{c||}{$p^{H}_{\perp}>30$} \\
\hline
& \textrm{PYTHIA} & \textrm{MC@NLO} & \textrm{PYTHIA} & \textrm{MC@NLO} \\
\hline
$N=0$ & 0.365 & 0.39 & 0.055 & 0.090 \\
\hline
$N=1$ & 0.335 & 0.345 & 0.40 & 0.465 \\
\hline
$N=2$ & 0.18 & 0.17 & 0.31 & 0.275 \\
\hline
$N=3$ & 0.080 & 0.060 & 0.15 & 0.105 \\
\hline
$N=4$ & 0.030 & 0.020 & 0.055 & 0.040 \\
\hline
$N>4$ & 0.010 & 0.015 & 0.030 & 0.025 \\
\hline
\end{tabular}
\caption{\label{NLOtab} Fraction of events with $N=0,1,2,3,4$ or more jets for inclusive Higgs boson production 
and Higgs boson production with $p^{H}_{\perp}>30$ GeV in PYTHIA and MC@NLO.}
\end{center}
\end{table}
\end{tiny}

\begin{figure}[h]
\begin{center} 
\includegraphics[width = 8cm,angle=0]{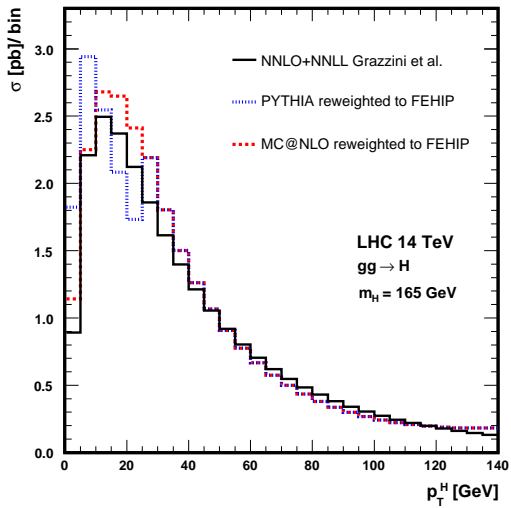}
\end{center}
\caption{\label{fig:ptresum}
The Higgs boson $p_{\perp}$ spectrum from a resummed calculation, MC@NLO reweighted and PYTHIA reweighted.}
\end{figure}

We next study what happens when we perform the reweighting at NNLO.  We use FEHIP to obtain these results.  
The inclusive NNLO cross section is 
given in Eq.(\ref{eq2}).  We include the reweighting of MC@NLO to the NNLO double differential distribution.  
The accepted cross sections for 
NNLO and the reweighted event generators are
\begin{equation} 
\sigma_{\rm acc}({\rm pb}) =
\left \{
\begin{array}{cl}
13.1, &\;\;\;{\rm R^{NNLO}(PYTHIA)};\\
14.9, &\;\;\;{\rm R^{NNLO}(MC@NLO)}; \\
14.9, &\;\;\;{\rm NNLO}.
\end{array}   
\right.
\label{NNLOcr}
\end{equation}
The acceptances are 
\begin{equation}
A =
\left \{
\begin{array}{cl}
0.47, &\;\;\;{\rm R^{NNLO}(PYTHIA)};\\
0.54, &\;\;\;{\rm R^{NNLO}(MC@NLO)}; \\
0.54, &\;\;\;{\rm NNLO}. 
\end{array}
\right.
\label{NNLOacc}
\end{equation}
We observe a much better agreement with the NNLO reweighting.  R(PYTHIA), R(MC@NLO) and the fixed order NNLO 
result all agree with the PYTHIA and MC@NLO acceptances within 6\%.  The NNLO result contains two partons in the final 
state, which gives a more realistic accounting of the QCD radiation.  It also contains the first radiative correction to the 
single parton $p_{\perp}$ spectrum.  The $p_{\perp}$ spectrum obtained at NNLO is in better agreement with MC@NLO, as seen in 
Fig.~\ref{fig:bin0a}.  A comparison of the $p_{\perp}$ spectrum from the reweighted generators with the resummed $p_{\perp}$ 
distribution of~\cite{ptresum} is presented in Fig.~\ref{fig:ptresum}.  There is good agreement between ${\rm R^{NNLO}(MC@NLO)}$ and the
resummed calculation.  ${\rm R^{NNLO}(PYTHIA)}$ agrees with the intermediate and large $p_{\perp}$ portion of the 
resummed distribution, while there is a slight discontinuity induced by the first bin reweighting in the low $p_{\perp}$ 
region.  We conclude that even in the presence of significant cuts on the jets in the final-state, 
the simple reweighting of the Higgs boson double differential distribution at NNLO describes the acceptances well.  In addition, 
since the NNLO result produces the correct 
normalization and contains drastically reduced scale dependences, we believe that reweighting MC@NLO with the fully differential 
NNLO result of FEHIP provides a very accurate prediction for the Higgs boson signal at the LHC.  

\begin{figure}[h]
\begin{center}
\includegraphics[width = 8cm,angle=0]{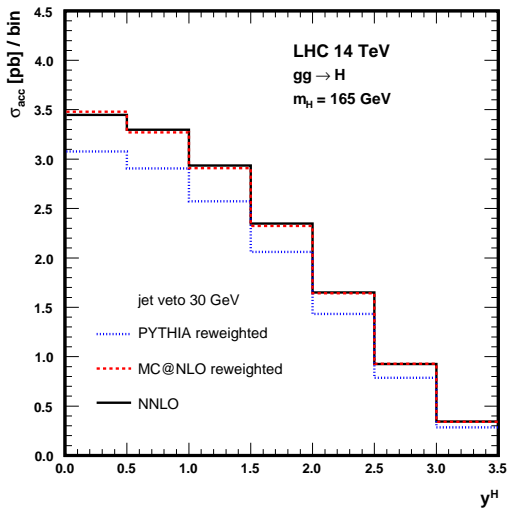}
\end{center}
\caption{\label{fig:yh_jv30} 
The Higgs boson rapidity distribution with a jet veto of 30 GeV applied for NNLO, MC@NLO reweighted and PYTHIA reweighted.}
\end{figure}

As a final check, we compute the rapidity distribution of the Higgs boson using FEHIP and the reweighted 
event generators.  The result is shown in fig.~\ref{fig:yh_jv30}.  We observe that imposing the jet veto 
maintains the matching of this distribution.

Motivated by the success of the NNLO reweighting procedure, we now allow the Higgs to decay and study predictions 
for the $pp \rightarrow H +X \rightarrow \gamma\gamma+X$ and $pp \rightarrow H \rightarrow W^+W^- \rightarrow l^-l^+ \bar{\nu}\nu$ 
channels.


\section{The di-photon channel}
\label{sec:twophoton}

We first use the reweighting procedure to compute the 
cross-sections for observables in the $pp \to H+X \to \gamma \gamma +X$ 
channel. The standard cuts on the two photons are
\begin{itemize}
\item  $p_\bot^{\gamma_1}> 40 {\rm \, GeV}$ and $p_\bot^{\gamma_2}> 25 {\rm \, GeV}$;
\item $\left|\eta^{\gamma_{1,2}}\right|<2.5$ ;
\item $E_{\mathrm{hadr}}< 15 {\rm \, GeV}$ in cones of size $R=0.4$ around each photon. 
\end{itemize}
The two-photon  channel is useful 
for additional  checks of the reweighting approach. Because 
the $H\to \gamma \gamma$ decay is included in the FEHIP program, 
we can compare observables computed with R(PYTHIA) and 
R(MC@NLO) directly with NNLO results.

We first compare the various results for the accepted cross-sections 
after applying  the standard cuts.  
We choose a Higgs mass  $m_H= 120 {\rm \, GeV}$ and set the 
renormalization and factorization scales to $\mu_R=\mu_F=m_H/2$. 
For the $H\to \gamma \gamma$ branching ratio we assume the value 
${\rm Br}(H \to \gamma \gamma)=0.002205$, which we obtain from HDECAY~\cite{Djouadi:1997yw}.  We find the following cross-sections 
for the di-photon signal:
$$
\begin{array}{||c|c|c|c|c|c||}
\hline
&\textrm{PYTHIA} & \textrm{MC@NLO} & R(\textrm{PYTHIA}) & R(\textrm{MC@NLO}) & \textrm{NNLO} \\ 
\hline
\sigma_{acc} [{\rm fb}]  
& 36.8 & 60.3  & 65.3 & 66.9   & 66.4 \\
\hline 
\end{array}
$$
The PYTHIA, MC@NLO and the NNLO results differ significantly,
reflecting again the large NLO and NNLO corrections. 
However, the reweighted cross-sections agree within $2.5\%$, 
and differ from the NNLO result only by $-1.7\,\%$ for R(PYTHIA)
and $+1.0\,\%$ for R(MC@NLO).

The effect of the cuts on the accepted 
cross-section  in the two-photon channel is rather insensitive to the choice of generator, and the 
reweighting procedure reproduces the NNLO results reliably. 
This is an expected result, since in the 
di-photon decay the experimental cuts do not resolve the 
structure of the hadronic system that recoils against the Higgs boson.

\begin{figure}[th]
\begin{center}
\includegraphics[width = 12cm,angle=0]{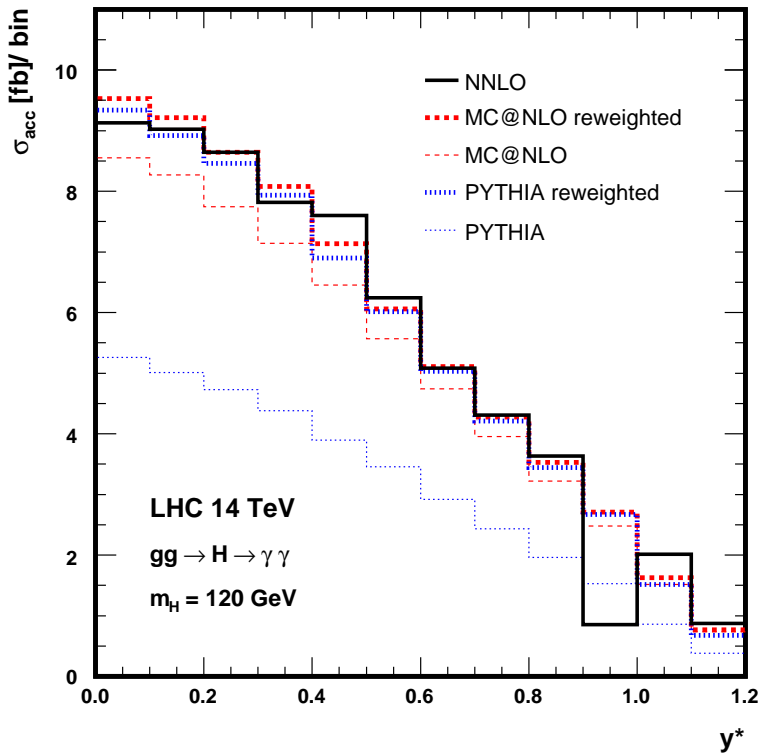}
\end{center}
\caption{\label{fig:ystar_pythia} 
The pseudorapidity difference $y^*$-distribution for di-photon 
events. We compare PYTHIA and MC@NLO with the reweighted generators and
NNLO.}
\end{figure}

We next compare the reweighted results and the 
NNLO predictions for more complicated observables in the di-photon 
channel. In Refs~\cite{lance1,fehip,fabian} the distribution of the 
pseudorapidity difference of the two-photons $y^* = 1/2 \left|
\eta^{\gamma_1} - \eta^{\gamma_2} \right|$ was
proposed as a discriminator
to separate the signal from the
prompt photon background. In Fig.~\ref{fig:ystar_pythia} we present 
the $y^*$-distribution for PYTHIA and MC@NLO, as well as for the reweighted generators
and at NNLO.  We observe  that the reweighted and the NNLO distributions agree 
reasonably well. At the boundary of the kinematic region 
$y^* \geq 0.96$, the NNLO distribution is non-monotonic. 
This kinematic region corresponds to a
vanishing  leading order cross-section, 
and the perturbative result must be resummed to all orders.
PYTHIA and MC@NLO do not suffer from this problem since 
parton showers perform such resummations.  The reweighted generators 
maintain the resummed behavior at the kinematic boundary and reproduce the fixed order 
result elsewhere.

\begin{figure}[th]
\begin{center}
\includegraphics[width = 12cm,angle=0]{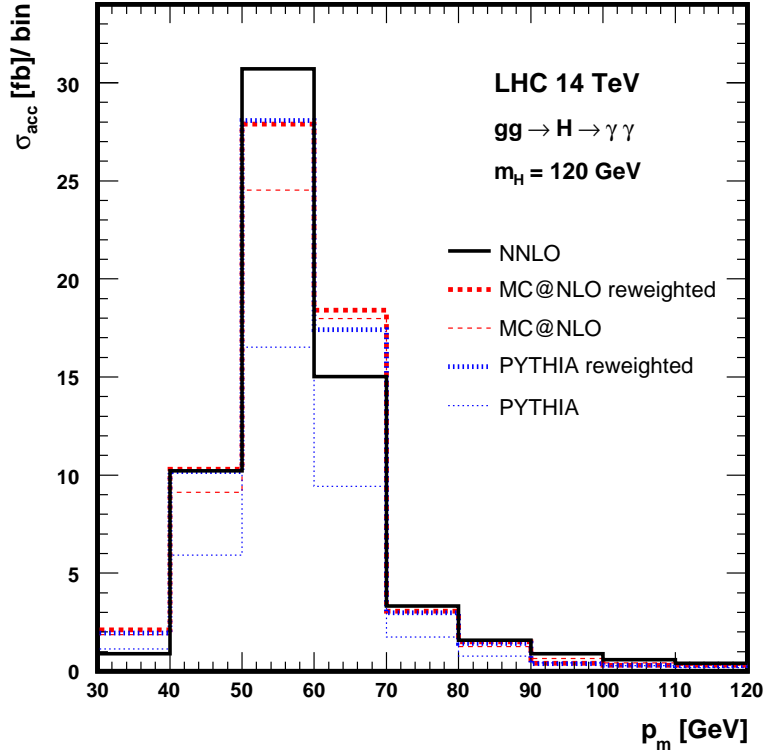}
\end{center}
\caption{\label{fig:p_m_pythia} 
The average transverse momentum $p_m$-distribution for the two
photons. We compare PYTHIA and MC@NLO with the reweighted generators and
NNLO.}
\end{figure}


\begin{figure}[h]
\begin{center}
\includegraphics[width = 10cm]{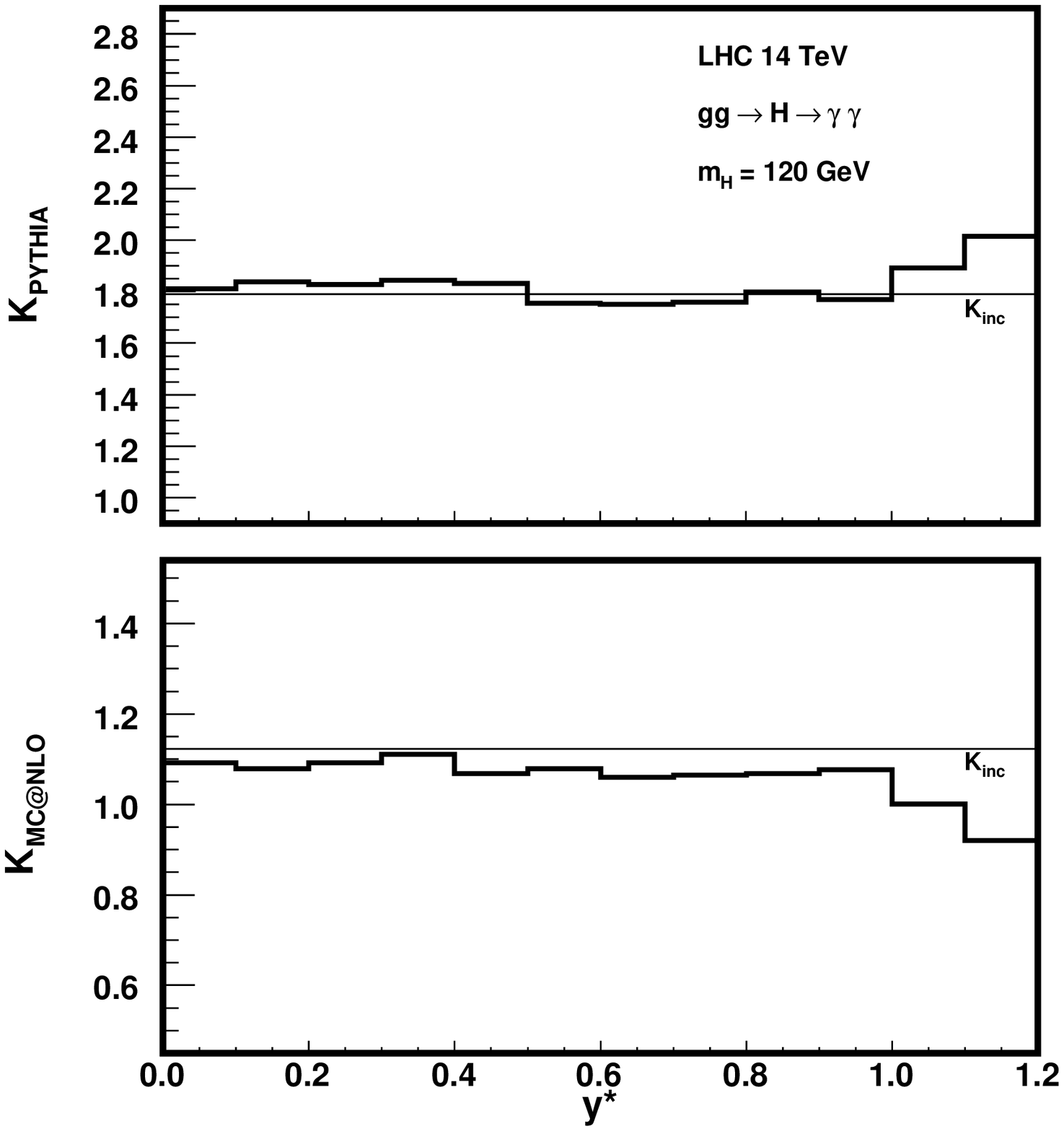}
\end{center}
\caption{\label{fig:ym_ratio} 
The effective $K$-factors as a function of $y^*$ for PYTHIA and MC@NLO.}
\end{figure}

We now study the average $p_\bot$ distribution of the 
two-photons, $p_m = 1/2 (p_\bot^{\gamma_1}+p_\bot^{\gamma_2})$~\cite{fehip,fabian}. 
At leading order in perturbation theory the cross-section is zero for 
$p_m > m_H/2$. The distribution at higher orders retains 
a characteristic peak at $p_m \sim m_H/2$. 
In Fig.~\ref{fig:p_m_pythia} we show the distribution for PYTHIA, 
MC@NLO, and FEHIP, and also after reweighting. We again find a very 
good agreement between the R(PYTHIA) and R(MC@NLO) results. 
The NNLO distribution agrees very well away from the peak at 
$p_m \sim m_H/2$. As expected, 
the NNLO result at the peak is substantially different because  
this region cannot be predicted accurately in fixed-order perturbation theory, 
and requires resummation.  The reweighted generators do a reasonably good job at 
maintaining the appropriate resummed behavior at the peak.

\begin{figure}[h]
\begin{center}
\includegraphics[width = 10cm]{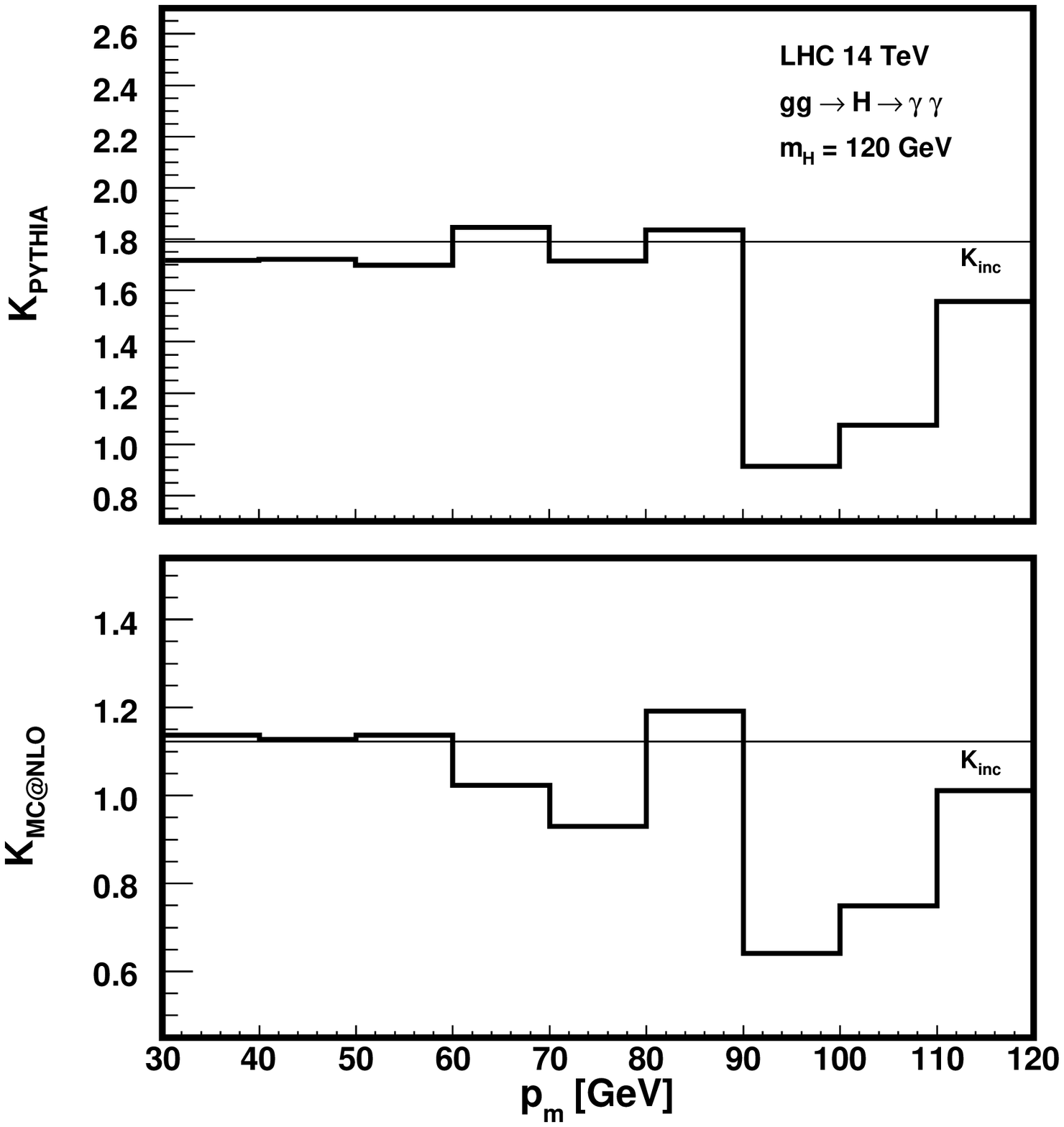}
\end{center}
\caption{\label{fig:pm_ratio} 
The effective $K$-factors as a function of the $p_m$ distribution for PYTHIA and MC@NLO.}
\end{figure}

The ratio of the di-photon cross-sections computed with the 
reweighted and the leading order generators is very similar to 
the NNLO $K$-factor for the total cross-section. The 
reweighting of the event generators with a constant factor could 
also yield realistic results for the di-photon cross-section after 
applying the standard cuts. 
It is interesting to investigate 
if a constant $K$-factor is also sufficient for reweighting differential 
distributions. We have already seen that this would not be satisfactory for the $p_\bot$ and rapidity distributions 
for the Higgs boson in Fig.~\ref{fig:kfac_pt_all}.  To investigate this,
we compute the effective $K$-factors for each bin of the $y^{*}$ and $p_m$ distributions,
\begin{equation}
K_G({\rm bin}) = \frac{\Delta \sigma^{R(G)}({\rm bin})}
{\Delta \sigma^{G}({\rm bin})}.
\end{equation}
In Fig.~\ref{fig:ym_ratio} we 
show that the effective $K$-factors in each bin of the $y^*$ 
distribution do not vary significantly from the inclusive $K$-factor. 
However, in the average photon $p_\bot$ distribution of Fig.~\ref{fig:pm_ratio}, 
the effective $K$-factors for high $p_m$ bins away from the peak are not uniform, although the 
large kinematic variations occur in bins with few events.

\section{The $W^+W^- \to l^+ l^- \nu \bar \nu$ channel}
\label{sec:Wchannel}

In this Section we study the reaction 
$pp \to H \to W^+W^- \to l^+l^- \nu\bar{\nu}$.
This signal can be distinguished from the main background process
$pp \to W^+W^-$
by applying  a jet veto, requiring a small opening angle between 
the two charged leptons in the transverse plane, 
and applying some additional kinematic cuts~\cite{Dittmar:1996ss,vetoth}.


In the remainder of this Section, 
we  present new results with the R(PYTHIA) generator in the 
$H\to W^+W^- \to l^+ l^- \nu \bar{\nu}$ decay channel. The observables that 
we consider  probe the momenta of the final-state leptons. The $W^+W^-$ decay 
chain is not yet implemented in FEHIP, and a comparison with NNLO is not 
possible. In addition, the Herwig event-generator, which is 
a basic component of MC@NLO, does not have an implementation of the same 
decay with full spin correlations. Therefore, we will not present 
any leptonic observables with R(MC@NLO). 

We first present a study of the $K$-factor for the $H \to W^+ W^- \to l^+l^-\nu \bar{\nu}$ 
channel.  Using R(PYTHIA), we can obtain a description of the Higgs boson rapidity 
distribution valid to NNLO, and examine the effect of this rapidity 
dependence.  The effective $K$-factor integrated over the whole region after all $H \rightarrow WW$ cuts
are applied is 2.098.  If we reweight PYTHIA to only the $p_{\perp}$ spectrum of 
the Higgs boson, we find an effective $K$-factor of 2.02.  
The $(p_{\perp},Y)$ dependent effective $K$-factor is 10\% lower than the fully 
inclusive $K$-factor of 2.28, while the $K$-factor coming from only reweighting to 
the $p_{\perp}$ distribution is 13\% lower than the inclusive $K$-factor, comparable with the 
results from~\cite{ptresum}.  The effect of the rapidity dependence is therefore less than 3\%.

We now employ R(PYTHIA) 
to study distribution shapes in $pp \to H \to W^+W^- \to l^+l^- \nu \bar \nu$.  We normalize 
each result to the integrated cross-section subject to the appropriate cuts. 
Since 
the distributions studied do not probe the hadronic radiation, we expect them to be very 
well described by R(PYTHIA).

\begin{figure}[h]
\begin{center}
\includegraphics[width = 8cm,angle=0]{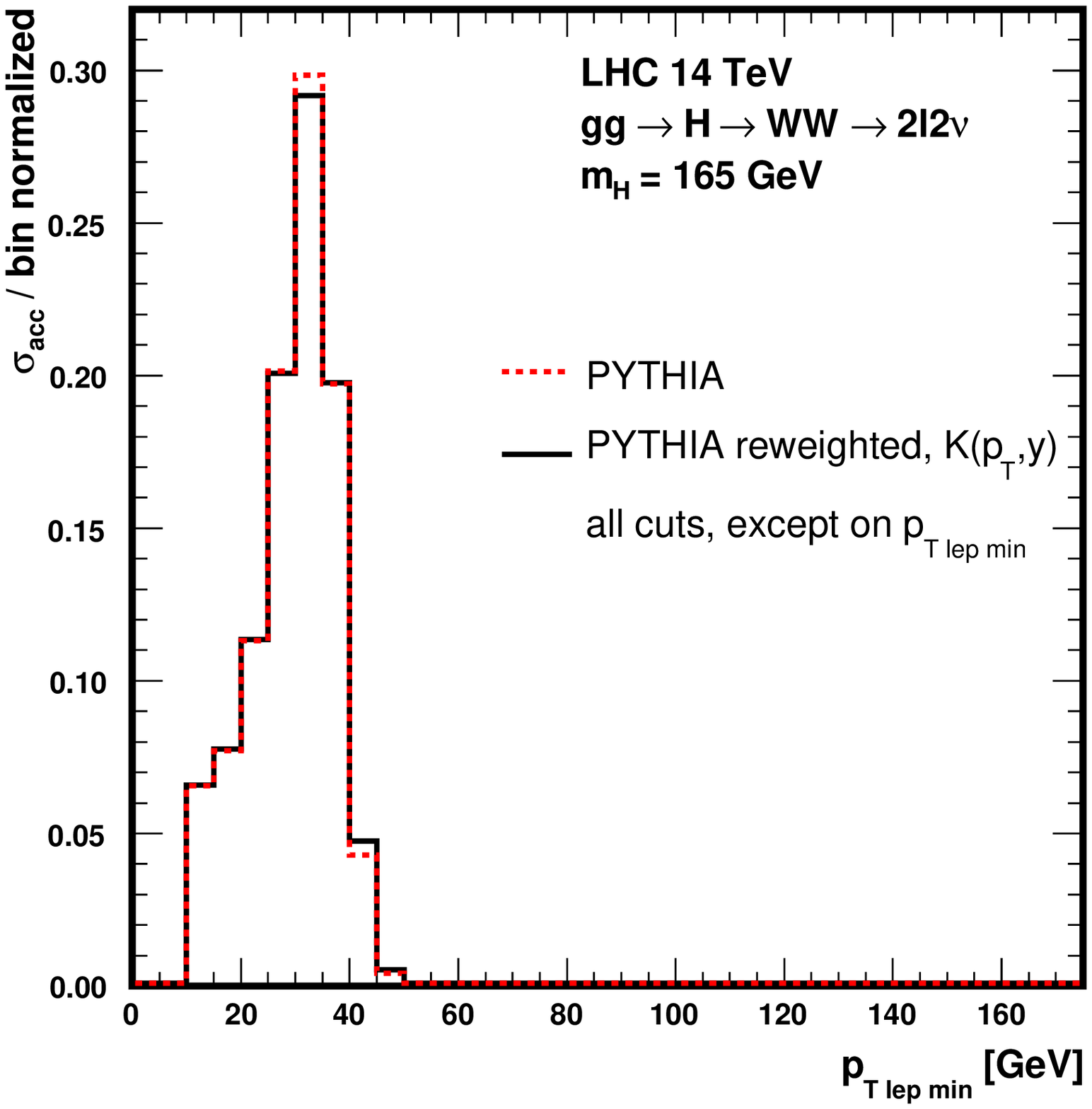}
\includegraphics[width = 8cm,angle=0]{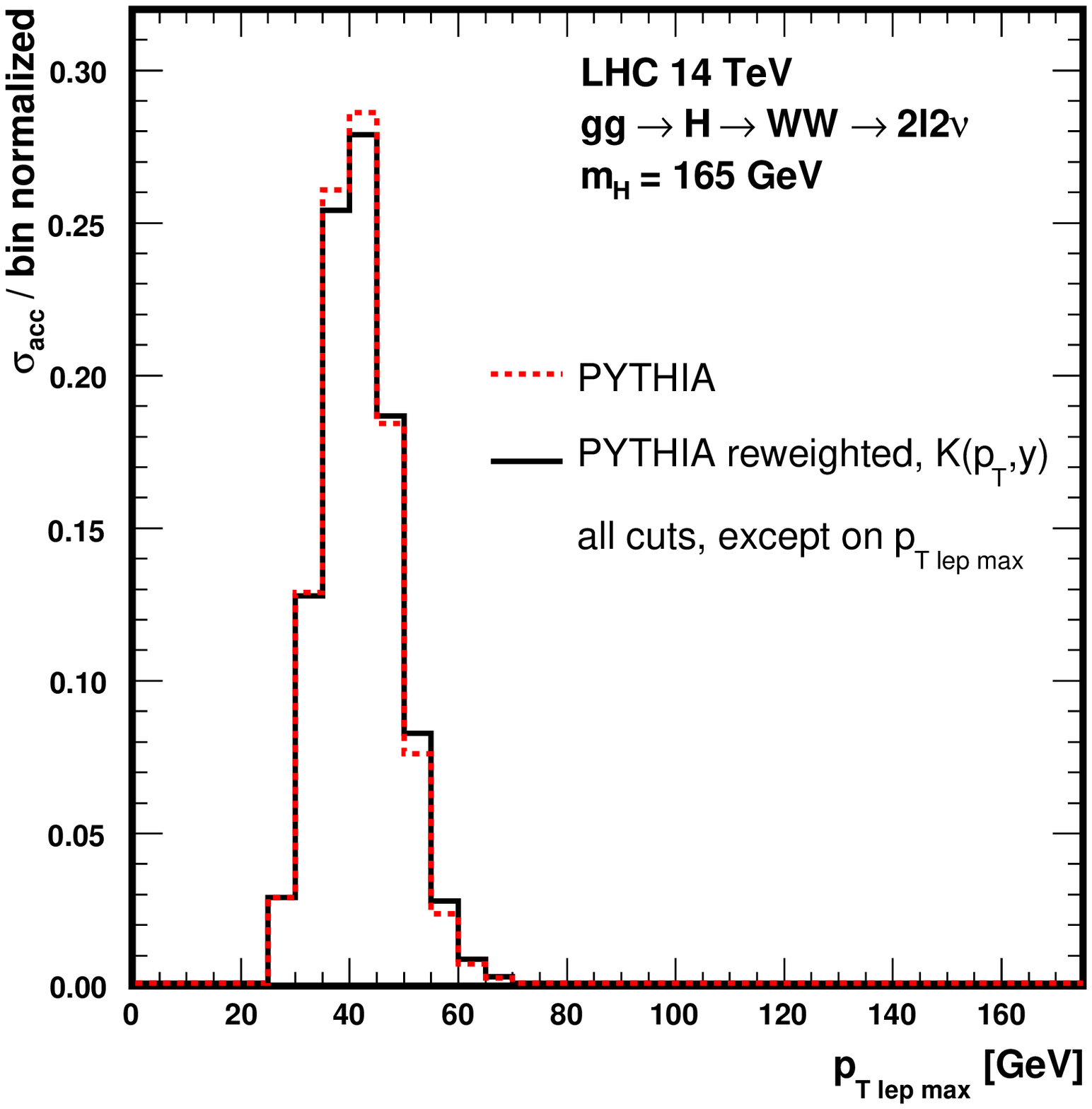}
\end{center}
\caption{\label{fig:ptlepton} 
The minimum (left) and maximum (right) 
transverse momentum of the two leptons computed with PYTHIA and R(PYTHIA).}
\end{figure}



In Fig~\ref{fig:ptlepton} we plot  the minimum and maximum transverse 
momentum distributions of the detected leptons for PYTHIA and
R(PYTHIA) events.  
These distributions are characteristic 
of the Higgs signal and can be used to discriminate from the background.  
%
%
We observe that the reweighting does not change the shape of distributions.
An application of a  constant $K$-factor would lead 
to the same results.  However, the appropriate $K$-factor is the 
effective one of 2.098 discussed at the beginning of this section, which 
is 10\% lower than the fully inclusive $K$-factor.


\section{Conclusions}
\label{sec:conclusions}

In this paper, we have presented a phenomenological approach to including 
NNLO corrections in event generators such as PYTHIA or MC@NLO.  Without an 
extension of the MC@NLO procedure to NNLO, this offers the best way of combining 
parton showering and hadronization with NNLO calculations.  We study this 
procedure for Higgs boson production at the LHC, since the fully differential 
NNLO calculation is available in the program FEHIP.  We reweight the Monte-Carlo events of PYTHIA and MC@NLO 
to match the bin-integrated NNLO double differential distribution in the Higgs $p_\bot$ and rapidity.  
We then study how well distributions of the Higgs boson decay products are predicted by this reweighting 
procedure.  We note that the $K$-factors that describe 
the reweighting of both PYTHIA and 
MC@NLO exhibit significant kinematic dependences, so that the use 
of a single constant $K$-factor may not be adequate.

We first study the reweighting procedure for the process $pp \to H+X$, without decays of the Higgs boson.  The 
$K$-factors that describe the reweighting of both PYTHIA and MC@NLO depend significantly on 
the Higgs boson transverse momentum, and non-negligibly on its rapidity. We test how well reweighting reproduces a 
fully consistent merging of fixed-order calculations with parton showering by reweighting 
PYTHIA at NLO and comparing to MC@NLO.  We find large discrepancies when a jet veto is imposed.  This 
indicates that the single parton emission present in the NLO calculation is insufficient to 
correctly describe cuts where the hadronic structure is probed.  The reweighting in the presence 
of a jet veto works much better at NNLO, where two partonic emissions are present in the 
final state.  This indicates the importance of extending perturbative calculations to NNLO in order 
to obtain a reasonable description of the additional radiation.

We then examine the decay channel $pp \to H \to \gamma\gamma$ with all relevant experimental 
cuts included.  We find that both the reweighted PYTHIA 
and the reweighted MC@NLO match very well the accepted cross-section as predicted by FEHIP.  We next study 
distributions that have been proposed to discriminate between the Higgs signal and the 
background.  Both R(PYTHIA) and R(MC@NLO) describe the kinematic distributions well.  They 
match the NNLO fixed-order result away from kinematic features, and exhibit the resummation 
present in the event generators near the kinematic boundaries.

We proceed to study the decay channel $pp \to H \to W^+W^- \to l^+l^- \nu\bar{\nu}$.  It is 
important to understand distributions in this channel at the LHC, since a direct reconstruction 
of the Higgs boson mass peak is not possible because of the two neutrinos in the final state.  We can not 
yet directly compare lepton distributions with the NNLO result, since FEHIP does not yet contain the 
decay $H \to W^+W^- \to l^+l^- \nu\bar{\nu}$. 
We study lepton and missing energy distributions using R(PYTHIA), assuming that it predicts the 
distribution shapes correctly.  Since Herwig does not yet contain spin correlations for this channel we do 
not present results for R(MC@NLO).  We find that the reweighting induces very small kinematic shifts.  
We study the effective $K$-factor for this channel after all cuts have been applied, and find that it is 
10\% smaller than the inclusive $K$-factor.  The effect of the rapidity dependence on this effective $K$-factor 
is small, about 3\%.

In summary, in this paper we study for the first time the detection efficiency for the Higgs boson at the LHC 
by reweighting parton 
shower Monte Carlo output to the fully differential 
Higgs boson cross section at NNLO in QCD.  Monte Carlo events from
PYTHIA and MC@NLO are normalized to the NNLO calculation of the Higgs boson rapidity and transverse momentum distributions.  
For Higgs boson events with low $p_{\perp}$, a constant $K$-factor is applied, to maintain the 
resummed shape present in the Monte Carlo simulations.  For the $H \rightarrow WW$ channel , where a jet veto is applied, 
we find a small difference of 3\% compared to a similar reweighting approach using only the transverse 
momentum spectrum.  We conclude that the effect of the NNLO Higgs boson rapidity dependence on LHC 
observables is now accurate to the percent level.  The dominant remaining theoretical uncertainties affecting 
the Higgs boson search at the LHC are: (1) the scale 
uncertainty arising from truncation of the 
perturbative expansion at NNLO; (2) the modeling of the low $p_{\perp}$ Higgs spectrum; (3) the theoretical 
uncertainties for the backgrounds to the Higgs signal.
We believe that the event reweighting studied here is a useful and accurate way of 
including higher order QCD calculations in Monte-Carlo event generators.  It allows us to 
determine the correct normalization at NNLO after all experimental cuts are included, 
incorporates the kinematic shifts induced by hard QCD radiation at higher orders, and maintains 
the resummation present in parton-shower Monte-Carlo programs near kinematic boundaries.  We believe that 
reweighting MC@NLO to match NNLO differential distributions gives a highly accurate description 
of the Higgs signal at the LHC.  We look forward to its application 
to other processes of phenomenological interest.

\bigskip\bigskip\bigskip
\noindent
{\Large\bf Acknowledgements}

\bigskip

\noindent
We thank M. Grazzini for assistance in obtaining the numerical results in~\cite{ptresum}.  We also 
thank S. Frixione and T. Sj\"{o}strand for useful comments.  K. M. is supported by the US Department of Energy  under contract 
DE-FG03-94ER-40833 and  
the Outstanding Junior Investigator Award DE-FG03-94ER-40833, and by the 
Alfred P. Sloan Foundation. F. P. is supported in part by
the University of Wisconsin Research Committee with funds granted by the
Wisconsin Alumni Research Foundation.


\end{document}